\documentclass[prd,aps]{revtex4}
\usepackage{amssymb,epsfig}

\begin{document}

\title{Geometrically motivated
hyperbolic coordinate conditions for numerical relativity: Analysis, issues and implementations}
\author{Carles Bona$^{1}$, Luis Lehner$^{2}$ and Carlos Palenzuela-Luque$^{2}$}

\affiliation{$1$ Departament de Fisica, Universitat de les Illes Balears, Palma de Mallorca,
Spain \\
$2$ Department of Physics and
Astronomy, Louisiana State University, Baton
Rouge, Louisiana 70803-4001, USA}

\begin{abstract}
We study the implications of adopting hyperbolic driver coordinate
conditions motivated by geometrical considerations. In particular,
conditions that minimize the rate of change of the metric
variables. We analyze the properties of the resulting system of
equations and their effect when implementing excision techniques.
We find that commonly used coordinate conditions lead to a
characteristic structure at the excision surface where some modes
are not of outflow-type with respect to any excision boundary
chosen inside the horizon. Thus, boundary conditions are required
for these modes. Unfortunately, the specification of these
conditions is a delicate issue as the outflow modes involve both
gauge and main variables. As an alternative to these driver
equations, we examine conditions derived from extremizing a scalar
constructed from Killing's equation and present specific numerical
examples.
\end{abstract}


\maketitle


\section{Introduction}
The choice of suitable coordinate conditions has certainly played
a major role in the understanding of solutions of Einstein
equations at the analytical level. From the early well posedness
result by Choquet-Bruhat \cite{choquet} to recent global existence
proofs~\cite{friedrich,christklain,rodlindblad} a judicious choice
of coordinates was key to yielding a tractable problem.

At the numerical level, coordinates play even a more crucial role as
an unfortunate choice might render the simulation useless despite
 much computational
effort. In fact, the optimal situation is one where coordinates
not only behave well (i.e. not forming coordinate singularities)
but also aid in the simulation. The latter refers to a choice of
coordinates that adapts to the problem at hand, making evident
(possibly approximate) symmetries that might be present.

A testimony of the importance of this subject has been the number
of works that have been devoted to it. From the early works of
York and Smarr \cite{yorksmarr,yorksources} which proposed coordinate
conditions through elliptic equations, to more recent works
which present alternatives to choosing coordinates that could
aid in the numerical simulation (see for
instance \cite{IBBH,GarGun99,ABDKPST03,LinSch03,BonPal04,BCP04})
considerable efforts have been invested towards defining useful
coordinate conditions.
In general, these conditions
seek to minimize suitably defined quantities with the hope that
these will, in turn, have a positive impact in the behavior of
 numerically evolved quantities.

The proposed coordinate conditions are usually given in algebraic
terms or through elliptic equations. The latter is expected when
conditions requiring stationarity of variables-- whose evolution
is determined by hyperbolic equations-- are imposed. When
attempting to use such conditions, one faces the problem of
dealing with a hyperbolic-elliptic system of equations which
require appropriate boundary conditions. Until recently, convenient
boundary conditions for the main variables in spacetimes involving
black holes were not sufficiently understood for generic
situations when singularity excision was
used \footnote{This picture has changed as consistent boundary
conditions have been developed applicable to cases where excision is employed
~\cite{dain,maxwell}.}. The complications associated with properly
defining the elliptic side of the problem at the analytical level
coupled to the additional ``extra
cost'' that solving these equations during the
evolution has spurred a number of efforts aiming to circumvent
both these issues. The idea has been to promote the elliptic
equations to hyperbolic ones through ``driver equations''
\cite{BDSSTW96}. These conditions aim to sidestep the cost issue
and the need to impose boundary conditions at possible excision
surfaces.

Conditions based on this strategy have been employed in a number
of works yielding much improved evolutions, most notably in BSSN
codes where specific coordinate conditions are obtained by
requiring the time variation of  the (trace of the) connection
variables be driven to zero (the so-called $\Gamma$-driver).
The hyperbolicity analysis of the BSSN system with
the $\Gamma-driver$ conditions augmented by suitable advection terms
has been presented in \cite{horstsarbach}. It is shown that the system
is strongly hyperbolic in this case, though unfortunately these augmented
coordinate conditions do not necessarily freeze the $\Gamma$ variables.
It is then unclear wether these augmented conditions will have the same
impact as the original ones in simulations and also if they share
similar hyperbolicity properties.
Therefore, there are a number of questions that remain open,
namely (i) What is the true impact of adopting
these `driver' conditions on the hyperbolic properties of the complete
(main variables plus gauge) system? (ii) Furthermore, are the
conditions obtained sufficiently flexible to guarantee desirable
properties such as to yield a convenient characteristic structure
at boundaries? (iii) How must one extend the knowledge gained in
the ``$\Gamma$-freezing conditions'' so as these can be thought as
truly geometric expressions not tied to particular variables in
the system --and hence useful to other formulations--? (iv) What
is the freedom in the implementation of these conditions and their
behavior in actual applications?

In the present work, we examine these questions both in the
theoretical and practical senses in order to draw conclusions
applicable to most metric based formulations of Einstein equations
by considering coordinate conditions motivated from possible
geometrical constructions. In particular, we concentrate on
conditions defined either at a given hypersurface (and its
embedding on the four-dimensional spacetime), or at the
four-dimensional spacetime level. The former results into a set of
elliptic equations which contains the well known minimal
distortion/strain condition for the shift vector and the maximal
conditions for the lapse while the latter gives hyperbolic
equations related, in a sense, to the harmonic coordinates.

With these conditions, we analyze the properties of the whole
system of equations (coordinate conditions plus Einstein
equations) where in the case of implementing  the elliptic
equations we promote them to hyperbolic ones via the ``driver''
approach. Additionally, we investigate possible difficulties that
can be encountered when employing these coordinate conditions in
conjunction with an excision strategy.

Our analysis mainly concerns 3+1 {\it
metric formulations}. That is, those based on the intrinsic metric
and extrinsic curvature of spacelike hypersurfaces defined by a
foliation of the spacetime. The equations governing the future
evolution of these variables are derived from the Einstein equations
in the spacetime of interest and are augmented by additional
variables and check whether that the resulting system, coupled to
the coordinate conditions, is at least strongly hyperbolic.

As we will see, in all cases one obtains a system with a characteristic
structure such that its eigenvectors couple gauge/coordinate variables to
the main variables. This has strong implications for the system, since:
\begin{itemize}
    \item In the cases where singularity excision is to be used, the
characteristics of the system must be such that they are completely
outflow towards the excision boundary. This condition is fulfilled
when, roughly $\beta^n > \alpha$ (with $\beta^n\equiv \beta^i
n_i$ the projection of the shift along the
spacelike unit normal to the excision
surface $n_i$  and $\alpha$ the lapse function). Since now the
coordinate conditions are dynamical,
extra care must be taken to monitor that it is fulfilled.
    \item
Since the characteristic modes of the system now mix
coordinate and main variables, if the condition above is not
satisfied, it is extremely difficult to provide boundary
conditions to the gauge functions consistently. This is to be
contrasted with the case where the elliptic conditions themselves
are employed. Here boundary conditions for the gauge variables
could be imposed so as to guarantee the outflow requirement is
satisfied.
\end{itemize}

Unfortunately, as we will describe in section III, commonly used
conditions fail to satisfy some important desirable conditions
which might have strong implications in numerical
implementations. We point out how this can be avoided and the cost
associated in doing so. Additionally, we analyze alternative
conditions derived from considering approximate symmetries in the
spacetime. This condition, called ``harmonic almost-Killing
equation'' coupled to Einstein equations gives rise to a well
behaved system where coordinates respond to (approximate)
symmetries. We present simulations to investigate their usefulness
within standard numerical relativity testbeds.

In order to carry out several analysis presented in this work,
it is necessary to adopt a given formulation of the equations. To this
end we consider the
strongly hyperbolic formulation presented in \cite{sarbachtiglio}
and the so-called Z4 formulation. The former can be regarded as an
`augmented' ADM formulation with the addition of first order
variables keeping the metric's gradient and suitable combination
of constraints to the right hand sides. The latter  is basically a
covariant extension of the Einstein field equations, obtained by
introducing a new four vector $Z_{\mu}$ which is defined by its
evolution equations. This way, the symmetrized covariant
derivatives of this four vector are added to the Einstein
Equations, that is
\begin{equation}\label{Z4}
  R_{\mu \nu} + \nabla_{\mu} Z_{\nu} + \nabla_{\nu} Z_{\mu} =
  8\; \pi\; (T_{\mu \nu} - \frac{1}{2}\;T\; g_{\mu \nu}).
\end{equation}
The solutions of the Einstein\'{}s solutions are recovered from
the extended set when $Z_{\mu}$ happens to be a Killing vector,
that is
\begin{equation}\label{lieZ}
    \nabla_{\mu} Z_{\nu} + \nabla_{\nu} Z_{\mu}~.
\end{equation}
For a generic spacetime, this happens just in the trivial case
$Z_\mu
= 0$, so true Einstein's solutions can be easily recognized.\\

\section{``Ideal coordinates''}

Before presenting different possibilities for adopting coordinates
we comment on what are arguably useful properties they should
satisfy. There exist several discussions on what constitute
requirements for good coordinate conditions (see, for instance,
\cite{yorksources,GarGun99,lehnerreview}); these are based on the
intuitive picture that useful coordinates, from the point of view
of a numerical implementation,  should:

\begin{itemize}
\item be free of artificial (coordinate) singularities; \item take
advantage of existing symmetries in the problem, whether
approximate or exact. In particular if the spacetime is
stationary, coordinate conditions should give rise to metric
components explicitly time independent; \item in the absence of
symmetries, they should minimize the rate of change of either the
metric or other appropriately defined geometrical quantities;
\item be 3-covariantly defined if possible (i.e. independent of
coordinate changes within a given hypersurface).
\end{itemize}

The points above are important at a fundamental level in that they
are hopefully satisfied irrespective of the formulation used or
the particular problem under study. To these, further requirements
might be added that refer more to specific applications like
coordinates having:

\begin{itemize}
    \item suitable behavior near singularities. For instance,
conditions yielding a convenient slicing of the spacetime. This
could range from those that avoid the singularities altogether
(like singularity avoiding conditions) to those that penetrate the
possible horizon (in the case where excision techniques are to be
applied). In the latter case, it important that the resulting
characteristic structure be such that all variables are outflowing
towards the excision region.
    \item appropriate asymptotic behavior
so that extraction of physically relevant quantities is
facilitated and/or, related coordinate speeds are bounded so as to
not have to deal with superluminal cases.
\end{itemize}

Finally, the conditions adopted must be such that within the
formulation of Einstein equations employed the well posedness of
the underlying problem to be treated is guaranteed as the choice
of coordinates is not decoupled from the issue of defining
a well posed problem. For instance,
the ADM formulation with analytically prescribed coordinate
(lapse and shift) conditions is weakly hyperbolic (hence yielding an ill-posed
problem) while with harmonic coordinates is symmetric hyperbolic.
For a more general discussion of hyperbolic formulations see for
instance \cite{reulareview}.

As mentioned, one of the main motivations when choosing
coordinate conditions is that they should not introduce spurious
``dynamics'' in the evolution of the system.
This motivation has lead throughout the
years to the introduction of different conditions. When attempting
to define such conditions an obvious difficulty is the need to do
so in an three-covariant way so as to decouple coordinate effects to the
true physical behavior of the spacetime. A way to do so was introduced
by Smarr and York \cite{yorksmarr} by constructing scalar quantities from
 the intrinsic and extrinsic curvatures of the spacetime and minimizing
their variation with respect to the lapse function and shift vector.

For instance, the `strain' scalar defined in analogy with fluid
dynamics, is defined as
\begin{equation}
  \partial_t \gamma_{ij} \equiv \Sigma_{ij} = 2~( \nabla_{(i} \beta_{j)} - \alpha
  K_{ij})
\end{equation}
which can be used to construct a positive definite Lagrangian
\begin{equation}\label{lagrangian3D}
  L \equiv \Sigma_{ij}~\Sigma^{ij} \, .
\end{equation}
This geometrical object is used to
construct a (non-negative) action which measures the
distortion or strain of a given hypersurface. By minimizing this
action with respect to the shift $\beta^i$,
\begin{equation}\label{variational}
   \frac{\delta S}{\delta \beta^i} = 0, ~~~ S \equiv \int L~\sqrt{g}~d^3x~.
\end{equation}
an elliptic equation (called ``minimal strain") is obtained for the
shift. When this
equation is fulfilled, the rate of change of a suitable norm of the
spatial metric will be minimized from one hypersurface to the next
one.

The minimal strain equation can be generalized by considering the
action of the densitized metric $\gamma^\lambda~\gamma_{ij}$ (with $\gamma = \det(\gamma_{ij})$),
obtaining this way what we will call the ``minimal densitized
strain". This equation can be written simply as
\begin{equation}\label{minimalstrain}
  \nabla_k [\Sigma^{ki} - \lambda~\gamma^{ki}~tr\Sigma] = \nabla_k [ \nabla^i \beta^k + \nabla^k \beta^i
                                        - 2~\lambda \gamma^{ki} \nabla_m  \beta^m
            - 2~\alpha~(K^{ij} - \lambda \gamma^{ki} trK)] =0
\end{equation}
so the minimal strain condition reduces to the choice $\lambda=0$.
Another interesting case, called ``minimal distortion" (in analogy
with a related notion of elasticity) is recovered for the choice $\lambda=1/3$.

These (elliptic) equations seem natural conditions for the shift
in Numerical Relativity applications, since they satisfy the
fundamental properties mentioned earlier. Notice that one could
have also opted to minimize the action with respect to the lapse
$\alpha$. This yields an algebraic condition for it
\cite{IBBH,GarGun99}, though it is ill-defined for time-symmetric
cases. An alternative strategy is to consider minimizing an scalar
defined by $\dot K_{ij} \dot K^{ij}$ with respect to the lapse.
This provides a fourth-order elliptic equation for it
\cite{dainpaper1}. We will refrain from considering it here as it
will introduce further complications either at the computational
cost level (if implementing the elliptic equation) or at a
conceptual level when promoting the equation to a hyperbolic
condition.

As mentioned, when considering these conditions within an initial
boundary value problem, assessing the well posedness becomes more
involved as the system becomes of elliptic-hyperbolic nature.
This, in particular, means the solution will strongly depend on
the boundary data specified. Until recently, the lack of a well
defined strategy to specify inner boundary conditions (say in the
case where black hole excision is adopted), coupled to the extra
cost associated with solving elliptic equations numerical induced
considerable activity towards employing related conditions within
a purely hyperbolic problem \footnote{These drawbacks are
presently not strong ones as reasonably well defined conditions
have been presented~\cite{dainetal} and efficient elliptic solvers
have been implemented
\cite{chop1,andersonmatzner,erickPHD,chop2,pfeiffer}.}. For these reasons,
it has become customary to define associated hyperbolic equations
to implement related  conditions. Parabolic conditions could also
be considered, but they certainly would not simplify the cost
issue (as their Courant-Friedrich-Levy condition scales
quadratically with the grid-spacing, and can be further regarded
as an inefficient method to solve the elliptic equation itself via
relaxation techniques). We will thus concentrate on hyperbolic
conditions and analyze the implications they carry.

\section{Hyperbolic Coordinate Conditions}
We will restrict our analysis to a large family of hyperbolic
coordinate conditions. Some of these are derived by simple
relations that have been employed in the past while others are
motivated by the minimization of geometrical quantities as
described above. In this latter case, we follow recent works
\cite{ABDKPST03,LinSch03}, in that we will approach the problem
here by adopting hyperbolic driver conditions to implement the
equations. However, as opposed to these works, we will require
that the coordinate conditions analyzed do indeed minimize the
desired elliptic equations. That is, we will neither assume that
considering other related equations having the same principal part
but differing in the lower order terms will yield similarly
behaved coordinates nor that suitable lower order terms can be
added so as to obtain first order conditions. Although these
assumptions can be, and have been, adopted in previous works, the
conditions thus obtained are not guaranteed to satisfy the
original sought-after geometrically motivated conditions.

To be specific, we will consider conditions that can be written as
\begin{eqnarray}\label{gaugeconditions}
    \partial_t \alpha &=& - \alpha^2 ~ Q  ~~, \\
    \partial_t \beta^i &=& - \alpha ~ Q^i
\end{eqnarray}
for suitably defined $\{Q,~Q^i\}$ which we regard as the gauge conditions.
These will be given by either algebraic or
differential equations relating the Q-quantities with the other
variables of the system, trying to fulfill as many
of the desired requirements described in the previous section as possible.

In what follows we consider separately three distinct cases which
refer to the way the fields $Q, Q^i$ are defined. We distinguish
cases with the name `algebraic' when $Q$ or $Q^i$ are directly
defined, `differential' when they obey evolution equations and
`semi-algebraic' in cases where an algebraic for one and a
differential for the other is considered.

\subsection{Algebraic gauge conditions}
One of the simplest choices is an algebraic relation
between the Q-quantities $\{Q,Q^i\}$ and the main variables of the
system. In this definition are included the general gauge
conditions proposed recently in \cite{LinSch03} and the
subfamilies discussed in many other works (e.g., \cite{GarGun99,BonPal04}).
The prototype of algebraic
gauge conditions is given by the harmonic coordinates, which were
introduced half a century ago to ensure the well posedness of
the Einstein Equations \cite{choquet,Bru62}. This is obvious as in this
case, the principal part of Einstein equations for all components reduce to
\begin{equation}\label{Z4_22}
   g^{\gamma \delta} \left( g_{\mu \nu} \right)_{,\gamma \delta} = l.o.
\end{equation}

Although the well posedness of the Cauchy problem is ensured this
way for the evolution system, this coordinate choice does not
fulfill ``a priori" many of the properties of an ideal gauge
condition in the absence of suitably defined sources or lower
order terms. In particular, the freezing of the metric components
in (almost) stationary spacetimes is by no means guaranteed.
Another delicate issue is that the shift so-defined is not a
three-vector and so need not reflect the symmetries in the
problem. This can be seen more clearly by translating the harmonic
coordinates condition to the 3+1 decomposition language:
\begin{eqnarray}\label{harmZ4_12}
   Q &=& -\frac{\beta^k}{\alpha}~ \partial_k \ln \alpha + trK  \\
   Q^i &=& -\frac{\beta^k}{\alpha}~ {\partial_k} \beta^i
   - \alpha~\gamma^{ki}~(\partial^j \gamma_{jk} - \partial_k \ln \sqrt{\gamma}
  - \partial_k ln \alpha )~. \nonumber
\end{eqnarray}
The $Q^i$ will not transform as a vector (except under linear
transformations), so neither $Q^i$ nor $\beta^i$ will be vectorial
quantities during the evolution.

\subsection{Semialgebraic gauge conditions}

We will refer to as semialgebraic gauge conditions those that,
keeping an algebraic relation for $Q$, allow for a differential
definition of the $Q^i$. This way there is enough freedom to fix
the shift with an exact geometric condition. Naturally, one could
have done the opposite, i.e. an algebraic relation to $Q^{i}$
while a differential one for $Q$ Since the former is what is most
commonly used in current applications and elliptic conditions for
the lapse, that minimize the rate of change of the extrinsic
curvature, is fourth-order we will concentrate on
algebraic/differential conditions for the lapse/shift.

For the lapse we propose a generalization of the harmonic
coordinates which includes the Bona-Masso lapse condition
\cite{BMSS95} and its slight modification presented in
\cite{ABDKPST03}, that can be written as:
\begin{equation}\label{BMlapse}
    Q = -a~ \frac{\beta^k}{\alpha} \partial_k ln \alpha + f(\alpha)~(trK - 2~\Theta)
\end{equation}
where $\Theta$ is added when considering the Z4 formulation of Einstein
equations, otherwise this term must be dropped.

The generalization consists on having added the parameter $a$ to
the above equations which determines wether the advection terms
are included ($a=1$) or not ($a=0$). The ``lapse speed'' (ie, the
speed of the eigenvectors associated to the lapse) will be fixed
by this parameter in combination with the free function
$f(\alpha)$. Notice also that the subfamily $a=1$ reduces to
several well studied cases depending on the expression for
$f\equiv f(\alpha)$: $f=0$ is the geodesic slicing, $f=1$ is the
time harmonic slicing, $f=2/\alpha$ is the ``1+log" and so on.
Additionally, the subfamily $a=0$ has been used successfully in
the evolution of single BH \cite{ABDKPST03}.

The rationale behind this generalization is that, as we will see
later, the characteristic structure of the coordinate conditions
adopted will have delicate, profound, differences depending on the
values these parameters take. In addition to this generalization for
the lapse condition, we consider a similar one to the shift condition
which we will define by a suitable hyperbolic driver which seeks
to satisfy the minimal distortion condition. We next revise how
this condition is to be defined and further generalize it to
include related options.

\subsubsection*{Approximate geometric shift}

An elliptic condition can be imposed in a dynamical way through a
parabolic or hyperbolic  ``evolution'' equation. The former can be
regarded as a standard relaxation way to obtain the solution of
the elliptic equation while the latter drives the solution
towards the desired one, in analogy with a damped
oscillator. The first approach was used in \cite{BDSSTW96} to
convert the minimal distortion elliptic equations into
time-dependent parabolic equations by means of the Hamilton-Jacobi
method, that is,
\begin{equation}\label{parabolicdriver}
  \partial_t \beta^i = \sigma ~\nabla_k [\Sigma^{ki}
                          - \frac{1}{3}~\gamma^{ki}~tr\Sigma] \, .
\end{equation}
The parameter $\sigma$ is characteristic of all the drivers, and
determines the dissipation strength employed so that the solutions
of the elliptic and the parabolic equations agree. For small
values of $\sigma$, the shift is expected to tend slowly to the
elliptic solution. At this point it is worth noticing that in the
fully dynamical case, it is not completely clear that equation
(\ref{parabolicdriver}) is actually a driver. The procedure is
inspired in simple elliptic equations, like for instance the
Laplace equation $\nabla^2 \phi=0$. In this simple case the
Hamilton-Jacobi method indeed provides a driver to the elliptic
equation. In the case of Einstein Equations however, this
conclusion is not immediate as the equations are highly coupled.
In a ``frozen variables'' approximation, where all main variables
are regarded as fixed, the driver condition does give rise to a
solution satisfying the elliptic equations (up to suitable
boundary conditions). In general, however, assessing this behavior
is considerably more delicate.

Nevertheless, current simulations indicate --at least for the
cases considered-- that the hyperbolic-driver conditions do give
rise to reasonably well behaved solutions \cite{ABDKPST03} as
judged by monitoring the approximate fulfillment of the original
elliptic condition that motivated the driver condition. These
simulations implement a driver in such a way that some of the
variables of the BSSN formalisms \cite{BauSha99,ShiNak95}, the
$\Gamma^i$, are frozen at late times in black hole evolutions.
Although they give rise to great improvement in the evolution of
single black holes an head-on collisions, it is not clear whether
they are successful due to the fact that they minimize
particularly delicate variables in the system \cite{huqlehner} or
due to their ``proximity'' (in a loose sense) to a minimal
distortion condition. If the latter is the reason, it would
indicate that this condition could benefit other formulations.
Unfortunately, to our knowledge, these conditions have been
employed in practical applications only in the BSSN-based codes.

In order to investigate the usefulness of the geometrically
motivated condition we consider it within the driver approach
generalized in the following way (Q3 equation):
\begin{equation}\label{Q3}
  (\partial_t  - b~\mathcal{L}_\beta) Q^i
  + \nabla_k [g~\alpha~(\Sigma^{ki}
             - \lambda~\gamma^{ki}~tr\Sigma)] = -\sigma~ Q^i \, .
\end{equation}
Let us discuss in detail the differences between the standard
gamma driver condition, as used in \cite{ABDKPST03}, and
(\ref{Q3}). First, a Lie term has been included in the Q3 equation
with a free constant $b$. Although this Lie term does not come
naturally from the ``driver", we will see later that it affects the
shift speed (that is, the speed of the eigenvectors associated to
the shift) and it will be required in order to fulfill other
requirements. The physical meaning, when $b=1$, would be that the
driver is not along the time lines but along the normal lines to
the space-like hypersurfaces.

The second difference is that we employ a covariant derivative
(with respect to the intrinsic metric of the hypersurface) in
(\ref{Q3}) which is dictated by the minimal distortion condition.
This additionally ensures the tensorial character of the equation,
and so both the $Q^i$ and the $\beta^i$ are now vectors, with the
corresponding advantages. Finally, the parameter $\lambda$ has
been kept in order to generalize the condition and adapt it to
other formalisms that do not use the conformal decomposition. This
way, it one can choose which densitized strain is going to be
minimized during the evolution.

\subsubsection{Characteristic Analysis}
In order to analyze the structure of the system with the
coordinate conditions adopted we must choose a particular
formulation. Here we employ the Z4 formalism though we have
checked that similar issues arise when employing the above
coordinate conditions in the formulation presented in
\cite{sarbachtiglio}.

The characteristic analysis of the gauge conditions
(\ref{BMlapse}, \ref{Q3}) with the Z4 formalism described in the
next section shows that there are three clearly separated ``gauge
cones''. We refer to them in this way to stress that these come
about due to the coordinate conditions considered. However, their
corresponding eigenvectors span {\it not just} the part of the
Hilbert space corresponding to the lapse, shift and derivatives of
this last one. Indeed, they have components both on the
coordinates and main variables sectors which will have delicate
consequences as we shall see later. These gauge cones can be
grouped into three distinct entities:
\begin{itemize}
  \item Lapse cone, which propagates with speed
    $ -\frac{a+1}{2} \beta_n \pm \sqrt{ f~\alpha^2 + (\frac{a-1}{2})^2~\beta_n^2 }$ .
  \item Transversal shift cone, which propagates with speed
  $ -\frac{b+1}{2} \beta_n \pm \sqrt{ g~\alpha^2 + (\frac{b-1}{2})^2~\beta_n^2 }$.
  \item Longitudinal shift cone, which propagates with speed
  $ -\frac{b+1}{2} \beta_n \pm \sqrt{ 2~g~(1-\lambda)~\alpha^2 + (\frac{b-1}{2})^2~\beta_n^2 }$.
\end{itemize}
An analysis of the associated eigenvectors both for the Z4 and the
formalism described in \cite{sarbachtiglio} reveals that the full
evolution system is strongly hyperbolic only if all the gauge
speeds are different one from each other and different from the
speed of light. Otherwise, there is a collapse of some of the
eigenvalues and there is not a complete basis of eigenvectors,
leading to a weakly hyperbolic system.

We thus see the need to introduce the Lie terms (controlled with
the parameters $\{a,b\}$) in equations  (\ref{BMlapse}, \ref{Q3})
which will provide sufficient flexibility to obtain a well behaved
system. For simplicity, let us focus on the condition for the
lapse ---the same discussion is also valid for the other gauge
cones---.
 If the Lie term is included ($a=1$), as the driver acts to minimize
the dynamics along the normal line, the associated speeds are
$-\beta \pm \sqrt{f}~\alpha$. This kind of structure allows for
inflow coordinates where $\beta_n > \sqrt{f}~\alpha$ and all the
lapse eigenvectors have negative speed. As mentioned, this
requirement is crucial near a black hole horizon, where a standard
practice is to excise the singularity by introducing an excision
boundary. Here it is not known which boundary conditions to define
and even how could be implemented if known~\footnote{If no
excision is employed this issue does not arise, see for
instance~\cite{ABCMMS98,HUEBNER}.}. Another problem of this
approach is the existence of so-called ``sonic points'' (in
analogy with fluid dynamics) where the speed is zero ($\beta_n =
\sqrt{f}~\alpha$). At these points there is a collapse of some of
the gauge eigenvectors with some standing modes, and the system is
weakly hyperbolic in one direction.

On the other side, if the Lie term is not included ($a=0$), the
lapse speeds are $ -\frac{1}{2} \beta_n \pm \sqrt{ f~\alpha^2 +
(\frac{\beta_n}{2})^2}$. In this case the driver is along the time
lines and the evolution system is always strongly hyperbolic; the
speeds/eigenspeeeds are such that only in the case $\alpha=0$
there could be a collapse of eigenvectors. However, some of these
eigenspeeds are such that, at an excision surface, will always
describe incoming modes (i.e. towards the computational domain).
This means that boundary conditions are to be specified for these
modes somehow. Unfortunately, as these modes couple coordinate and
main evolution variables, one has to worry about how to provide suitable
boundary conditions to the gauge functions and carry the evolution
of the main variables without providing boundary conditions to
them. This is a delicate problem in itself. Intuitively, one expects
that boundary conditions are only required for the gauge functions; however,
some of the main variables themselves depend on the gauge functions also
(the extrinsic curvature). Hence, the issue of separating the gauge
dependent component of the incoming modes must be clarified before proceeding
this way. To do so requires considering constraint preserving boundary
conditions which might be further
complicated by the fact that the characteristic structure need
not be constant along the excision surface.

Summarizing, there is a tension between trying to obtain a minimizing
prescription and ensure both strong hyperbolicity of the system and
that any suitably defined excision boundary is of outflow type.
In hindsight it could be argued that this is a consequence of having
tried to `get away' without solving an elliptic equation --which does
require boundary conditions at all boundaries-- and solely deal with
a hyperbolic equation where no boundary is required at the excision surface.

It is then clear that the options are: (i) to stay at the elliptic (or related
parabolic)
level for the coordinate conditions; (ii) give up the symmetry seeking
approach through driver conditions (at least in the problematic regions
by suitably modifying the equations or by adding convenient lower order
terms to the equations \cite{frans} ) or (iii) consider a new set of options
that aim to resolve the conflicts.

\subsection{Almost-Stationary Motions: the Q4}
An appealing alternative is to consider conditions derived by
minimizing some suitably defined spacetime scalars. As it has been
recently pointed out in \cite{BCP04}, the harmonic almost-Killing
equation (HAKE)
\begin{equation}\label{HAKE}
   \nabla_\mu [\xi^{(\mu;\nu)} - \frac{1}{2} (\nabla \cdot \xi)~g^{\mu\nu}] = 0
\end{equation}
is a generalization of the Killing equation $\xi^{(\mu;\nu)}=0$
whose solution space includes also the affine Killing vectors and,
of course, its subfamily the Homothetic Killing vectors. For this
reason, the covariant conservation law (\ref{HAKE}) can provide a
precise definition of the concept of approximate Killing vectors
as solutions of the HAKE equation (\ref{HAKE}). This equation can be obtained
from the standard variational principle (\ref{variational}) with a
Lagrangian $L$ given by
\begin{equation}\label{lagrangianQ4}
   L = \xi_{(\rho;\sigma)}~\xi^{(\rho;\sigma)} - \frac{1}{2} (\nabla \cdot
   \xi)^2~~.
\end{equation}
Since the Lagrangian is non positive it is not possible
 to guarantee that extremizing the action will provide a
solution that minimizes it. However, by suitably adding damping
terms, the hope is that will indeed be the case. In such a case,
the HAKE equation can be of great utility as a coordinate
condition, because it is not only well adapted to the stationary
spacetimes (a Killing vectors is a solution) but also ``minimizes"
the deviation from the stationary regime. In spacetimes with some
(quasi) symmetry, it is expected that the congruence of time lines
of the observers will be aligned during the evolution with the
time (almost) Killing vector, avoiding this way spurious time
dependence due to an unfortunate choice of coordinates.

The physical meaning can be better understood by considering the
adapted coordinates $\xi =\partial_t$, where now $\Sigma_{\mu \nu}
\equiv \mathcal{L}_\xi g_{\mu \nu} = \partial_t g_{\mu \nu}$. The
4D Lagrangian (\ref{lagrangianQ4}) can be written as
\begin{equation}\label{lagrangianQ4-adapted}
   L = \Sigma_{\mu \nu}~\Sigma^{\mu \nu}
     - \frac{1}{2} g_{\mu \nu}~(\Sigma_{\gamma \delta}~g^{\gamma \delta}) .
\end{equation}
which can be reinterpreted as a four-dimensional generalization of
the positive definite  3D lagrangian (\ref{lagrangian3D}). Following the
analogy, the HAKE equation (\ref{HAKE}) can be seen as the 4D
generalization of the minimal densitized strain
(\ref{minimalstrain}). The main difference is that, since the
HAKE equation considers also the time component of the spacetime, the structure of the
resulting system is not elliptic anymore but hyperbolic.

In the Z4 context there are Z-terms that must be included in the
HAKE (\ref{HAKE}) in order to get a well posed problem. With these
terms, the conservation law (\ref{HAKE}) can be written in
different ways, like for instance
\begin{equation}
\label{HAKE1}
 \nabla_\nu [\frac{1}{\sqrt{g}}~{\cal L}_{\xi} ( \sqrt{g}~g^{\mu\nu})] =
 2~g^{\mu \nu}~{\cal L}_{\xi} Z_\nu ~~,
 \end{equation}
 or, in adapted coordinates,
\begin{equation} \label{HAKE2}
 g^{\sigma \rho}~(\partial_t {\Gamma^{\mu}}_{\sigma \rho}) +
 2~g^{\mu \nu}~\partial_t Z_\nu = 0 ~~.
\end{equation}
Equation (\ref{HAKE1}) shows explicitly the tensorial character of
the gauge condition, while the equation (\ref{HAKE2}) points out
its relation with the harmonic coordinates, i.e. $g^{\sigma \rho}~
{\Gamma^{\mu}}_{\sigma \rho} = 0$. This gauge condition, which
will be called Q4, is the closest to fulfill all the requirements;
not only the shift but also the lapse is well adapted to
stationary spacetimes, and if there is only an approximated
symmetry, the coordinates are expected to adapt in order to
minimize the rate of change of the metric. An additional property
is that as a result of their construction, the gauge conditions
obtained are also defined in a covariant way.

The ambiguity of including or not the Lie terms in the
Q-equations, introduced in the Q3 gauge, is not present here,
where there is no choice: the Lie terms are actually there. As a
consequence, the lack of strong hyperbolicity at sonic points
appears again in the gauge cones, as it will be shown in the next
section.

\section{The evolution system: Z4 formalism + Q4 gauge}

In order to study the hyperbolicity of the gauge conditions we
have to consider them within the context of a specific formalism in
order to get a closed set of equations that will constitute the
evolution system. For concreteness we adopt the Z4 formalism, but
similar results can be obtained with other formulations.

Here the Z4 formalism and the Q4 gauge will be written down as an
evolution system of (second order in space and first order in
time) equations by means of the 3+1 decomposition. The
characteristic structure of a fully first order version of this
evolution system will be analyzed in detail, as well as how to
pass from a second order system (in space) to a first order one
without altering the structure of the eigenvectors.

\subsection{The Formalism : the (first order) Z4 system}

The four-dimensional equations (\ref{Z4}) can be written, by using
the 3+1 decomposition, in the equivalent form \cite{BLPZ03}:
\begin{eqnarray}
\label{dtgamma}
  (\partial_t -{\cal L}_{\beta})~ \gamma_{ij} &=& - 2~\alpha~K_{ij}
\\
\label{dtK}
   (\partial_t - {\cal L}_{\beta})~K_{ij} &=& -\nabla_i\alpha_j
    + \alpha~   [R_{ij}
    + \nabla_i Z_j+\nabla_j Z_i
    - 2~K^2_{ij}+ ({\rm tr} K - 2~ \Theta)~K_{ij}
    - S_{ij}+\frac{1}{2}~(trS - \tau)~\gamma_{ij}]
\\
\label{dtTheta} (\partial_t -{\cal L}_{\beta})~\Theta &=&
\frac{\alpha}{2}~
 [R + 2~ \nabla_k Z^k + (trK - 2~ \Theta)~ trK
 -tr(K^2)  - 2~ Z^k {\alpha}_k/ \alpha - 2~\tau]
\\
\label{dtZ}
 (\partial_t -{\cal L}_{\beta})~Z_i &=& \alpha~ [\nabla_j~({K_i}^j
  -{\delta_i}^j \, trK) + \partial_i \Theta
 - 2~ {K_i}^j~ Z_j  -  \Theta~ {{\alpha}_i/ \alpha} - S_i] \:\:\:.
\end{eqnarray}
In order to convert the equations (\ref{dtgamma}-\ref{dtZ}) into a fully
first order system, the spatial derivatives of the lapse, the shift
and the intrinsic metric must be introduced as new independent quantities, that is
\begin{equation}\label{space_derivatives_metric}
 A_i~\equiv~ \partial_i \ln \alpha,
 ~~{B_{k}}^i~\equiv~ \partial_k \beta^{i},
 ~~D_{kij}~\equiv~\frac{1}{2}\;\partial_k \gamma_{ij}
\end{equation}
and substituted everywhere. The evolution equations for these
additional quantities can be computed easily taking the time
derivative of the definition (\ref{space_derivatives_metric}) and
permuting the time and spatial derivatives. Due to the commutativity
of second spatial derivatives, we can add without any change in the
solution space the constraints $C_{ki} =
\partial_k A_i -
\partial_i A_k$, ${C_{lk}}^i = \partial_l {B_k}^i - \partial_k
{B_l}^i$ and $C_{klij} = \partial_l D_{kij} -
\partial_k D_{lij}$ with free parameters $\{ c_a ~c_b ~c_d\}$ to the evolution
equations of the lapse, shift and intrinsic metric respectively.
It the evolution equations for the metric components are defined
in a general way as
\begin{eqnarray}\label{gaugeconditions}
    \partial_t \alpha = - \alpha^2 ~ Q  ~~,~~
    \partial_t \beta^i = - \alpha ~ Q^i ~~,~~
    \partial_t \gamma_{ij} = - 2~\alpha ~ Q
\end{eqnarray}
then the evolution of their spatial derivatives, with the addition
of the ordering constraints, would be
\begin{eqnarray}
   \partial_t A_{i} &+& \partial_i [ \alpha~Q ] - c_a~\beta^l~C_{li}=0
\\
   \partial_t {B_k}^i &+& \partial_k [ \alpha~Q^i] -
    c_b~\beta^l~{C_{lk}}^i = 0
\\
  \partial_t D_{kij} &+& \partial_k [ \alpha~Q_{ij} ] - c_d~\beta^l~C_{lkij}= 0
\end{eqnarray}
Here there is a delicate point; if one wants to preserve the same
eigenvectors when passing from the second order system (in space)
to the first order one, the choice $\{ c_a~c_b~c_d\} = \{ 1
~1~1\}$ is compulsory. Since we are interested on the physical
solutions, which should not depend on the order of the (spatial
derivatives of the) equations, this will be our choice from now
on.

With this choice, a first order version of the Z4 formalism can be
written as a system of balance laws:
\begin{eqnarray}
\label{dtA1}
\partial_t A_{i} &+& \partial_l [ -\beta^l~A_{i} +
        {\delta^l}_i~(\alpha~Q + \beta^m~A_{m} ) ] = {B_i}^l~A_{l} - {B_l}^l~A_{i}
\\
\label{dtB1}
 \partial_t {B_k}^i &+& \partial_l [ -\beta^l~{B_k}^i +
        {\delta^l}_k~(\alpha~Q^i + \beta^m~{{B_m}^i} ) ] = {B_l}^i~{B_k}^l - {B_l}^l~{B_k}^i
\\
\label{dtD1}
 \partial_t D_{kij} &+& \partial_l [ -\beta^l~D_{kij} +
        {\delta^l}_k~(\alpha~Q_{ij} + \beta^m~D_{mij} ) ] = {B_k}^l~D_{lij} - {B_l}^l~D_{kij}
\\
\label{dtK1}
\partial_t K_{ij} &+&  \partial_k [- \beta^k~K_{ij} + \alpha\; {\lambda^k}_{ij}~] =
S(K_{ij})
 \\
\label{dtTheta1}
\partial_t \Theta &+&  \partial_k [- \beta^k~\Theta + \alpha\; (D^k - E^k - Z^k)~]
 = S(\Theta)
\\
\label{dtZ1}
\partial_t Z_i &+& \partial_k [-\beta^k~Z_i + \alpha~
 \{-{K^k}_i + {\delta^k}_i (trK - \Theta) \}~]
 = S(Z_i)
\end{eqnarray}
where
\begin{eqnarray}\label{flux_K}
    {\lambda^k}_{ij} = {D^k}_{ij}
     - \frac{1}{2}~(1 + \xi)\; ({D_{ij}}^k + {D_{ji}}^k)
    &+& \frac{1}{2} {\delta^k}_i\; [A_j + D_j - (1 - \xi)\; E_j -2~Z_j] \\
    &+& \frac{1}{2} {\delta^k}_j\; [A_i + D_i - (1 - \xi)\; E_i -2~Z_i] ~~,
    \nonumber
  \end{eqnarray}
being $D_i\equiv {D_{ik}}^k$ and $E_i \equiv {D^k}_{ki}$. The
non-zero source terms can be found in the Appendix.

\subsection{The Q4 gauge} %

The gauge condition can be written as a set of evolution equations for some
gauge quantities by means of the 3+1 decomposition, using either
the covariant conservation law (\ref{HAKE1}) or the non-vectorial
``standing" equation (\ref{HAKE2}). Of course, these equations
(and their 3+1 forms) are completely equivalent, so one could be
recovered from the other without any problem at the second order
(in spatial derivatives) level. At the first order level this
transition is not always so transparent when the ordering
constraints $C_{ki},~{C_{lk}}^i$ and $C_{klij}$ are included; this
is the reason to start from the appropriate version of the HAKE equation
from the very beginning, at the four-dimensional level, to write
then the most convenient 3+1 gauge equations.

The 3+1 form of the conservation law (\ref{HAKE1}) provides
directly evolution equations for the tensor quantities $\{ Q,~Q^i
\}$, and it can be useful to take advantage of the symmetries of
the problem. For instance, in spherical coordinates the vector
$Q^i$ is in general $Q^i = (Q^r, Q^\theta, Q^\phi)$. If the
problem is also spherically symmetric, then only $Q^r$ and
$\beta^r$ would have a non-trivial evolution equation, as opposed
to what happens either with the harmonic coordinates
(\ref{harmZ4_12}) or the non-vectorial standing version
(\ref{HAKE2}). Notice that the semialgebraic Q3 (\ref{Q3}) has
also this vectorial character.

On the other side, the 3+1 form of the ``standing" version
(\ref{HAKE2}) reduces directly to evolution equations for some
combinations of variables which follow an ODE, so they are
directly standing modes of the system. This version is more
convenient in general cases without symmetries in order to write
the system in fully first order. The resulting standing modes,
combinations of the Q-quantities with other variables of the
system, hold always, so these eigenvectors are the same in both
second and first order versions. The ``standing" version
(\ref{HAKE2}) can be written, by means of the 3+1 decomposition,
as
\begin{eqnarray}\label{HAKEP}
  \partial_t P &\equiv& \partial_t [ \alpha~(Q - trK + 2~\Theta) + \beta^j ~A_j ] =
   -2~\alpha^2~K_{ij}~(Q^{ij} - \gamma^{ij}~Q) - 2~\alpha~Q^j~(A_j + Z_j)
\\ \label{HAKEPv}
  \partial_t P^i &\equiv& \partial_t [ \alpha~Q^i + \beta^j~{B_j}^i +
    \alpha^2~(2~E^i - D^i - A^i + 2~Z^i)] = \\
   && 2~\alpha~Q^j~(\alpha~{K_j}^i - {B_j}^i)
    + 2~\alpha^3~(Q^{jk} - \gamma^{jk}~Q)~{\Gamma^i}_{jk}
    + 4~\alpha^3~(Q^{ij} - \gamma^{ij}~Q)~Z_j
    - \alpha~Q^i~[\alpha~(Q-trK) + \beta^j~A_j] \nonumber
\end{eqnarray}
where the standing P-quantities have been defined. From these
equations it is easy to see that the principal part is just the time
derivative of the harmonic conditions (\ref{harmZ4_12}).

As it was discussed previously, equations
(\ref{HAKEP}-\ref{HAKEPv}) admit many different solutions. A
convenient way to enforce the precise desired solution without
unfavorably affecting the characteristic structure of an hyperbolic system was
introduced in \cite{BFHR99}. The method consists in adding a
source damping term that damps the solution to the desired one. In
our case it would be:
\begin{eqnarray}\label{dampingQ}
  \partial_t Q &=& ... - \sigma~(Q - \eta~trQ)
\\ \label{dampingQv}
  \partial_t Q^i &=& ... - \sigma~Q^i
\end{eqnarray}
where the dots stand now for all the original terms. Since only
one vector can be constructed just by contracting the
$\Sigma_{\mu\nu}$ tensor with the normal lines, there can not be
any ambiguity on the damping term for the equation of $Q^i$.
However, the two different scalars $Q$ and $trQ$ can be
constructed from $\Sigma_{\mu\nu}$, so all the combinations are
included in the damping terms in (\ref{dampingQ}). Two special
cases arise here:
\begin{itemize}
    \item The first one would correspond to the choice
$\eta=0$; all the Q-quantities are driven to zero, so that one
tries to minimize the rate of change of all the metric components.
It is the default case, most suitable in physical situations in
which we expect a stationary regime to be reached asymptotically.
    \item The second case would correspond to the choice
$\eta=1$; the lapse equation (\ref{HAKEP}) is driven to the
solution $Q=trQ$ instead of $Q=0$. This way, although the shift
equation is still used for minimizing the intrinsic metric, the
lapse just tries to follow the singularity avoidant condition
$\partial_t (\alpha/\sqrt{\gamma})=0$. This can be useful in all
cases in which singularity avoidance is required. Note that a
stronger singularity avoidance behavior is expected when $\eta >
1$.
\end{itemize}

\subsection{Characteristic structure} %

The evolution system have the following $54$ independent variables
\begin{equation}\label{u1}
 u ~ = ~ \{\alpha,~\beta^{i}, ~\gamma_{ij},~ K_{ij},~\Theta,~Z_k,
       ~A_i, ~{B_j}^i,~ D_{kij},~ P, ~P^i \}
\end{equation}
where the $\{ Q, Q^i \}$ can be written in the equations as
function of the $\{ P, P^i \}$ and other variables. The system is
strongly hyperbolic if all the eigenvalues are real with a
complete base of eigenvectors for any arbitrary direction $n^k$ and
the symmetrizer can be shown to be smooth. We have not looked into
this, though the analysis would follow the lines of those presented
in \cite{nagyreulaortiz,horstsarbach} where the
condition has been shown to hold.

The analysis of the eingenvalue/eigenvector structure
is more clear when the quantities (and the modes) are
decomposed by projecting them into this specific direction. That
way, for instance, a vector $T_i$ would be separated into its
longitudinal part $T_n \equiv T_k~n^k$ and its transversal
components $T_a \equiv T_i - T_n~n_a$. From now on we will use the
indices $\{a,b,c,d \}$ for the transverse components and $n$ for
the projection along $n^k$. Using this notation, the list of the
gauge-independent eigenvectors can be written as:
\begin{itemize}
    \item Standing modes: there are 10 eigenfields corresponding
    to the metric components with speed $v=0$
    \begin{equation}
       [\alpha]~,~~~[\beta^i]~,~~~[\gamma_{ij}] \, .
    \end{equation}

    \item Normal modes: there $20$ transversal components of the first order
    derivatives, propagating with speed $v=-\beta^n$:
    \begin{equation}
       [D_{cij}]~ ,~~~ [{A_c}]~, ~~[{B_c}^i] \, .
    \end{equation} 

    \item Transverse light  cone:  there
    are 6 new independent eigenvectors propagating with
    light speed $v = -\beta_n \pm \alpha$ that allow to
    recover $\{K_{ab},D_{nab}\}$, that is,
    \begin{equation}
     {L_{ab}}^{(\pm)} = [~K_{ab} - \frac{1}{2~\alpha} (B_{ab} + B_{ba})~]
           \pm [~D_{nab} - \frac{(1+ \xi)}{2}~ (D_{abn} + D_{ban})~] \, .
    \end{equation}

    \item Mixed light cone: there are 4 new independent
    eigenvectors propagating with light speed $v = -\beta_n \pm \alpha$
    that allow to recover $\{K_{na},Z_{a}\}$, that is,
    \begin{equation}
     {L_{na}}^{(\pm)} = [K_{na}] \pm \frac{1}{2}~
     [~A_a + {D_{ac}}^c + (\xi-1)~{D^c}_{ca} - \xi~D_{ann} - 2~Z_a~] \, .
    \end{equation}

    \item Energy cone:  there are 2 new independent
    eigenvectors propagating with light speed $v = -\beta_n \pm \alpha$ that allow to
    recover $\{ \Theta,~ Z_n\}$, that is,
    \begin{equation}
      E^{(\pm)} = [~\Theta - \frac{1}{\alpha} {B_c}^c~]
      \pm [~{D_{nc}}^c - {D^c}_{cn} - Z_n~]~~.
    \end{equation}

\end{itemize}

Several comments are in order here. The eigenvectors are simple
due to the choice $\{c_a~c_b~c_d \} = \{1 ~1 ~1 \}$. Other cases,
like the trivial one $\{c_a ~c_b ~c_d \} = \{0 ~0 ~0 \}$, are
considerably more complicated and do not have a complete basis of
eigenvectors at the sonic points. Notice also that, up to here,
the results are independent on the gauge condition. The remainder
of the characteristic structure is dictated by the choice of
coordinate conditions; in our case it is given by
\begin{itemize}

    \item Lapse sector:  The standing eigenvector $[P]$, plus the cone
    spanned by the 2 new independent eigenvectors
    propagating with light speed $v = -\beta_n \pm \alpha$ that allow to
    recover $\{K_{nn}, A_{n}\}$, that is,
    \begin{equation}
     {G}^{(\pm)} = [ P ] + (\alpha \mp \beta_n)~
     [~(trK - 2~\Theta) \pm A_n~]
    \end{equation}

    \item Transversal shift sector: The standing eigenvector $[P_a]$,
    plus the cones spanned by the 4 new independent eigenvectors
    propagating with light speed $v = -\beta_n \pm \alpha$ that allow to
    recover $\{B_{na},D_{nna}\}$, that is,
    \begin{equation}
     {S_{a}}^{(\pm)} = [ P_a ] + (\alpha \mp \beta_n)~
     [~\alpha~( A_a + D_a - 2~E_a -2~Z_a) \pm (B_{na} + B_{an})~]
    \end{equation}

    \item Longitudinal shift sector: The standing eigenvector $[P_n]$, plus
    the cone spanned by the 2 new independent eigenvectors
    propagating with light speed $v = -\beta_n \pm \alpha$ that allow to
    recover $\{B_{nn},D_{nnn}\}$, that is,
    \begin{equation}
     {S_{n}}^{(\pm)} = [ P_n ] - (\alpha \mp \beta_n)~
     [~\alpha~D_n \pm (\alpha~trK - trB)~] \,.
    \end{equation}
\end{itemize}

Note that at the sonic points ($\mid\beta_n \mid = \alpha$) one of
the sign choices in the former equations coincides with one of the
standing eigenfields $\{P,P_i\}$. Then, there is not a complete
basis of eigenvectors and the system is just weakly hyperbolic
there, as it happens with the Q3 gauge. However, these sonic
points can be found only in the tachyonic coordinates regions,
where $\beta^2 \ge \alpha^2$. Moreover, there will be missing
eigenvectors there only for the particular directions in which
$\mid\beta^k n_k\mid = \alpha$. This is considerably less severe
than the failure to achieve strong hyperbolicity for generic
directions in the full computational domain. This sonic points
issue arises also in the hydrodynamical equations and is often
dealt with by a small amount of dissipation~\cite{kreiss}. In the
next section we will check numerically the behavior of the system.

\section{Numerical Results}

The previous gauge conditions will be tested in different periodic
spacetimes which have been suggested \cite{Mexicotests} as
standard test-beds for Numerical Relativity codes. The numerical
algorithm used is the standard method of lines \cite{MoL} with
centered second order discretizations of spatial derivatives and
third order Runge--Kutta to evolve in time. We will focus on three
different tests; the `robust stability' test, in order to check
the well posedness of the formalism. The gauge waves, in order to
see the effect of the different gauge conditions in spacetimes
with a time-like Killing vector. And the Gowdy waves, where there
is no such time-like Killing vector. Usually we will compare the
results of the (first order) Z4 formalism either with the harmonic
coordinates (the evolution system will be called Z4harm) or the Q4
condition (the evolution system will be called ZQ4). The results
for the zero shift case, which are identical in few cases to the
Z4harm as we will show later, were already presented in
\cite{BLPZ04}.

\subsection{Robust stability}

Let us consider a small perturbation of Minkowski space-time which
is generated by providing random initial data for every dynamical
field in the system. The level of the random noise must be small
enough to make sure that, as long as the implementation is stable,
fields will remain at the linear regime even for a hundred
crossing times (the time that a light ray will take to cross the
longest way along the numerical domain). This test is designed to
experimentally assess the hyperbolicity of the evolution system by
exciting high frequency modes and observing the overall behavior
of the solution. As higher frequencies are allowed in the problem,
for a strongly/symmetric hyperbolic systems the solution should be
well behaved while this is not the case with weakly hyperbolic
systems.

The results of this test  around the standard flat space-time
$\beta^i=0$ are already well known from the analytical analysis
for both the evolution systems Z4harm and ZQ4. Since both of
them are strongly hyperbolic around $\beta^i=0$, all the norms
remain constant during the simulation, decreasing slightly due to
the inherent dissipation of the numerical scheme (no additional
artificial dissipation has been added in the simulations). However, it can be
useful to study numerically what happens at the sonic point
$\beta^x = \alpha$, where the ZQ4 is strongly hyperbolic for all
the directions but one, and check whether the ZQ4 system leads to
a convergent solution. In order to study this scenario, we define
both $\alpha$ and $\beta^x$ being $1$ plus a small random perturbation.

\begin{figure}[h]
\begin{center}
\epsfxsize=8cm
\hspace{8mm}
\epsfbox{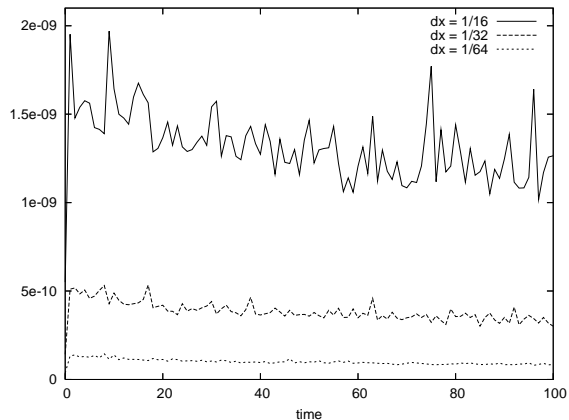}
\end{center}
\caption{Maximum norm of the $trK$ for the harmonic gauge around
the sonic point $\beta^x \approx \alpha \approx 1$ for three
different resolutions. The slope of the norms remains constant
independently of the resolution as expected on strongly hyperbolic
systems. The simulation are performed in a cube of length $L=1$
with 16, 32 and 64 points respectively. The time step is
$dt=0.25~dx$ and no artificial dissipation has been added.
}\label{robusttrKharm}
\end{figure}

\begin{figure}[h]
\begin{center}
\epsfxsize=8cm
\hspace{8mm}
\epsfbox{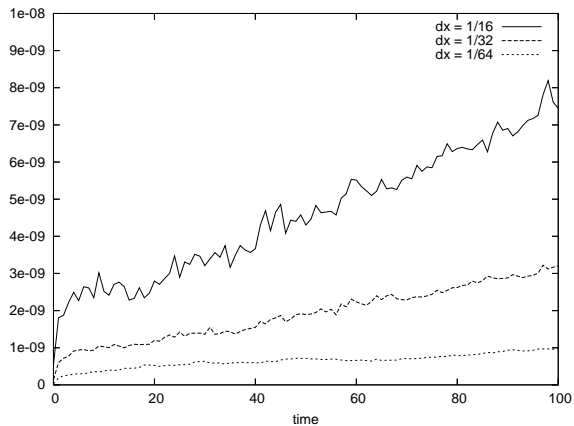}
\end{center}
\caption{The same plot that in Fig.~\ref{robusttrKharm} but for
the Q4 gauge. The slope of the norms, although they are growing in all the
simulations, decreases as the resolution increases, approaching
to the (constant) exact solution. This suggests that the system is
well posed even at the sonic points. }\label{robusttrKQ4}
\end{figure}

In order to see the expected behavior, we have plotted first in
Fig.~\ref{robusttrKharm} the norm of $trK$ of the Z4harm evolution
system around the sonic point for three different space
resolutions. Notice that, although we are displaying one
scalar quantity, the same behavior is observed by all the other
norms. As it is expected in strongly hyperbolic systems, as resolution
is increased the numerical solution either should not grow
or its growth should be lesser. Note
also that the same kind of behavior is shown for both the Z4harm
and the ZQ4 evolution systems around $\beta=0$.

A similar plot is presented in Fig.~\ref{robusttrKQ4} for the ZQ4
evolution system, again for three different space resolutions.
Although some of the variables ($trK$ in the plot) show a growing
norm, its slope decreases with resolution. So, although the
profiles are different from the standard case shown in
Fig.~\ref{robusttrKharm}, the observed behavior is consistent wit
a stable implementation, suggesting that the ZQ4 evolution system
is not ill posed at the sonic points.

\subsection{The gauge waves}

We now consider again the Minkowski metric written in a
non-trivial coordinates, obtained by performing a general
conformal transformation to the t-x coordinates, that is,
\begin{equation}\label{minkwaves}
   ds^2 = H^2(t,x)~(-dt^2 + dx^2) + dy^2 + dz^2~~.
\end{equation}
Propagation along the x axis can be simulated by considering a
dependence like $H(t,x)=h(x-t)$, so the exact time evolution (in
these coordinates) will be just the ``shifted" initial profile.
We will use here a periodic smooth profile, like a sine wave
\begin{equation}\label{sinus}
   h(t,x)= 1 - A~\sin(\frac{2~\pi~x}{d})
\end{equation}
where $d$ is the size of the x domain and $A$ is the amplitude of
the wave. Additionally, we will take advantage of the periodicity
of the initial profile to use periodic boundary conditions with
$d=1$.

\begin{figure}[h]
\begin{center}
\epsfxsize=8cm
\hspace{8mm}
\epsfbox{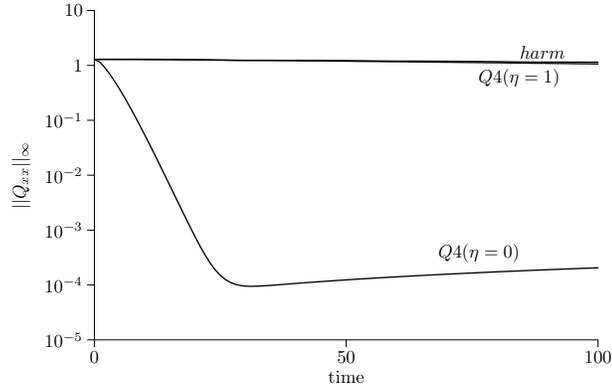}
\end{center}
\caption{Norms of the non-trivial strain component $Q_{xx}$ for
both the harmonic and the Q4 gauges with different values of the
second damping parameter $\eta$. While the harmonic and the Q4
gauges with $\eta=1$ show a similar non-freezing behavior, the Q4
with $\eta=0$ actually minimizes the strain, driving the system to
a stationary state. The initial amplitude of the gauge wave is
$A=0.1$ and the simulations are performed in a channel of
$50\times5\times5$ points with length $L=1$ in the longest
direction. The time step is again $dt = 0.25~dx$ and in this case
some (small) Kreiss-Oliger artificial dissipation has been added
in order to kill the high-frequency modes. }\label{gwavestrQnorms}
\end{figure}

\begin{figure}[h]
\begin{center}
\epsfxsize=8cm
\hspace{8mm}
\epsfbox{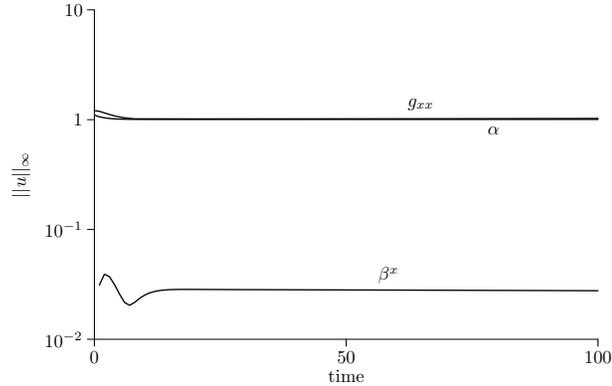}
\end{center}
\caption{The norms of the metric components for the ZQ4 evolution
system with $\eta=0$. After few crossing times all of them remain
almost constant, implying a very small value of its time
derivatives, the Q-quantities.}\label{gwavesmetricnorms}
\end{figure}

In Fig.~\ref{gwavestrQnorms} the norm of the strain $Q_{xx}$ is
shown for both the Z4harm and the ZQ4 evolution systems. The
evolution of the Z4harm is the same as the zero shift case
described in \cite{BLPZ04}. In the ZQ4 case we have plotted two
different cases, corresponding to different damping coefficients.
The first one is with $\eta=1$, so the time lines are driven to
the condition $Q=trQ$. As it can be seen in the plot, the result
is very similar to the harmonic case. The second case corresponds
to the choice $\eta=0$, where the time lines are driven to get
aligned with one of the Minkowski time-like Killing vectors. The
result shows the desired behavior; the observers behave in a way
in which the metric components are explicit stationary, as it can
be checked in Fig.~\ref{gwavesmetricnorms}.

\begin{figure}[h]
\begin{center}
\epsfxsize=8cm
\hspace{8mm}
\epsfbox{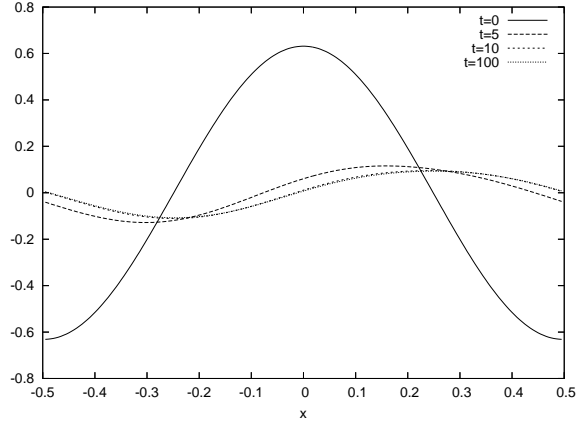}
\end{center}
\caption{The extrinsic curvature $K_{xx}$ in the x direction at
different times for the same simulation that in
Fig.~\ref{gwavesmetricnorms}. After 10 crossing times there are
not many changes in the profile.}\label{gwaveskxx}
\end{figure}

\begin{figure}[h]
\begin{center}
\epsfxsize=8cm
\hspace{8mm}
\epsfbox{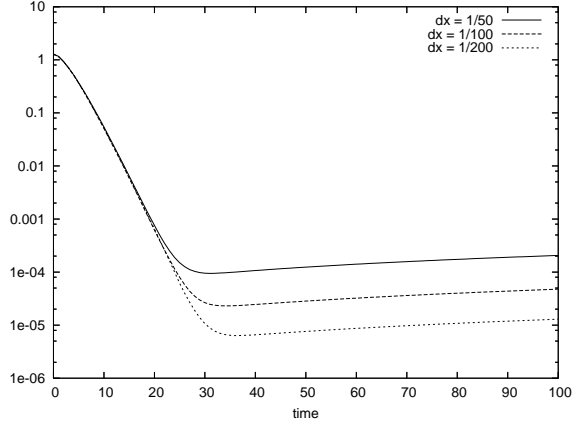}
\end{center}
\caption{The non-trivial component $Q_{xx}$ is plotted for
different resolutions ($dx=1/50, ~dx=1/100,~dx=1/200$) with the
Z4Q with $\eta=0$. All plots match during the transient decaying
stage, until some minimum is reached. This minimum value can be
seen to converge at a second order rate to
$Q_{xx}=0$.}\label{gwavesconvergence}
\end{figure}

Different snapshots of the (non-trivial component of the)
extrinsic curvature $K_{xx}$, in Fig.~\ref{gwaveskxx}, shows that
the evolution is almost frozen between 10 and 100 crossing times.
Finally, a convergence test for the variable $Q_{xx}$ is performed
in Fig.~\ref{gwavesconvergence}. The solution displays a decaying
behavior, in which all resolutions match, until an
asymptotic stage is reached (after around 30 crossing times),
where the plots clearly converge to $Q_{xx}=0$.

\subsection{The Gowdy waves}
Let us consider now the Gowdy spacetime, which
describes a space-time containing plane polarized gravitational
waves. The line element can be written as
\begin{equation}\label{gowdy_line}
  {\rm d}s^2 = t^{-1/2}\, e^{{\cal Q}/2}\,(-{\rm d}t^2 + {\rm d}z^2)
  + t\,(e^{\cal P}\, {\rm d}x^2 + e^{-{\cal P}}\, {\rm d}y^2)
\end{equation}
where the quantities ${\cal Q}$ and ${\cal P}$ are functions of $t$ and $z$ only
and periodic in $z$, so that (\ref{gowdy_line}) can be implemented
with periodic boundary conditions. Following \cite{Mexicotests},
we will choose the particular case
\begin{eqnarray}
\label{function_P}
  {\cal P} &=& J_0 (2 \pi t)\; \cos(2 \pi z)
\\
\label{fucntion_L}
  {\cal Q} &=&  \pi J_0 (2 \pi) J_1 (2 \pi)
   -2 \pi t J_0 (2 \pi t) J_1 (2 \pi t) \cos^2(2 \pi z)
\nonumber \\
  &+& 2 \pi^2 t^2 [{J_0}^2 (2 \pi t)  + {J_1}^2 (2 \pi t)
  - {J_0}^2 (2 \pi)  - {J_1}^2 (2 \pi)]
\nonumber \\
\end{eqnarray}
so that it is clear that the lapse function
\begin{equation}\label{gowdy_lapse}
  \alpha = t^{-1/4}\; e^{{\cal Q}/4}
\end{equation}
is constant everywhere at any time $t_0$ at which $J_0(2\pi t_0)$
vanishes. In \cite{Mexicotests} the initial slice $t=t_0$ was
chosen for the simulation of the collapse, where $2 \pi t_0$ is
the 20-th root of the Bessel function $J_0$, i.e. $t_0\; \simeq
\;9.88$.

Let us now perform the following time coordinate transformation
\begin{equation}\label{gowdy_time}
  t~=~t_0\;e^{-\tau / \tau_0}, ~~ \tau_0~=~t_0^{3/4} e^{{\cal Q}(t_0)/4}~\simeq~472\;,
\end{equation}
so that the expanding line element (\ref{gowdy_line}) is seen in
the new time coordinate $\tau$ as collapsing towards the $t=0$
singularity, which is approached only in the limit
$\tau\rightarrow\infty$. Notice that this singularity avoiding
time coordinate $\tau$ is not the proper time nor it does coincide
with the number of crossing times, due to the collapse of the
lapse.

\begin{figure}[h]
\begin{center}
\epsfxsize=8cm \hspace{8mm} \epsfbox{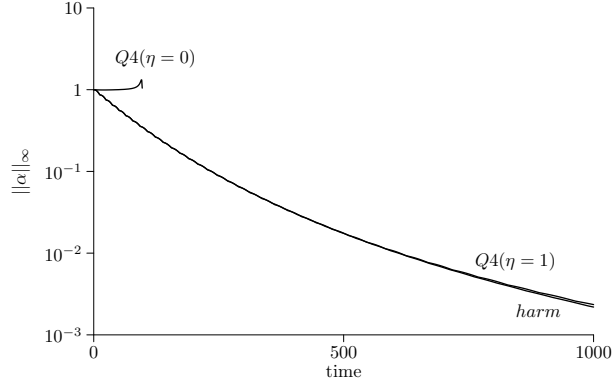}
\end{center}
\caption{Maximum norm of the lapse for both the harmonic and the
Q4 gauges. After 100 crossing times the Q4 gauge with $\eta=0$
gets too close to the singularity and crashes, while the other
cases continue evolving until 1000 crossing times without problem.
The simulations are performed in a channel of $5\times
5\times 50$ points with length $L=1$ in the longest direction. The
time step is again $dt = 0.25~dx$ and some amount of Kreiss-Oliger
dissipation has been added. }\label{gowdyalpnorm}
\end{figure}

\begin{figure}[h]
\begin{center}
\epsfxsize=8cm \hspace{8mm} \epsfbox{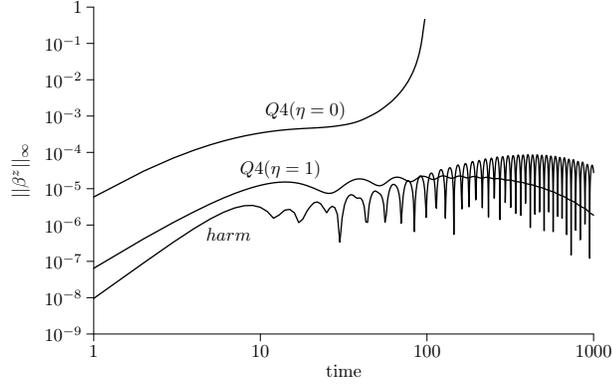}
\end{center}
\caption{Maximum norm of the non-trivial component of the shift
for the same simulation than in Fig.~\ref{gowdyalpnorm}. The time
scale has been changed from the linear to the logarithmic type in
order to clarify the plot. Note that the Q4 gauge with $\eta=0$
leads to a monotonic growing of the norm of the shift. In the other cases, we
can see the strong oscillations of the harmonic case as contrasted
with the smooth behavior of the Q4 gauge with
$\eta=1$.}\label{gowdybxnorm}
\end{figure}

As in the gauge waves test, the simulation is performed for the
Z4harm and the ZQ4 evolution system with different values of
$\eta$. The maximum norm of the lapse $\alpha$ is plotted in
Fig.~\ref{gowdyalpnorm}, showing two different kinds of
behavior\begin{itemize}
    \item Singularity avoiding behavior. This is indicated by the
collapse of the lapse, which can clearly be seen both in the
harmonic gauge and in the ZQ4 case with $\eta=1$. This behavior is
very similar to the zero shift case already described in
\cite{BLPZ04}. We can see in Fig.~\ref{gowdybxnorm} that the rate
of change of the shift is much slower in the ZQ4 case with
$\eta=1$ than in the harmonic case, where strong time oscillations
appear. This shows the freezing effect of the Q4 gauge, when
compared with harmonic simulations, even for singularity avoiding
choices of the gauge damping parameters.
    \item Lapse freezing behavior. This is indicated by the
absence of lapse collapse in the ZQ4 case with $\eta=0$. Since
the metric is collapsing, one gets close to the singularity in a
finite amount of coordinate time and the code crashes. When
translated in terms of proper time, however, all the simulations
arrive approximately to the same point. We can see in
Fig.~\ref{gowdybxnorm} a sharp increase in (the norm of) the
shift, which is trying to freeze the collapse by increasing the
observers outgoing speed.
\end{itemize}
It is worth to note here that this qualitative difference in the
numerical simulations is triggered by the choice of the second
damping parameter $\eta$, without affecting the principal part of
the original HAKE equation.

\begin{figure}[h]
\begin{center}
\epsfxsize=8cm
\hspace{8mm}
\epsfbox{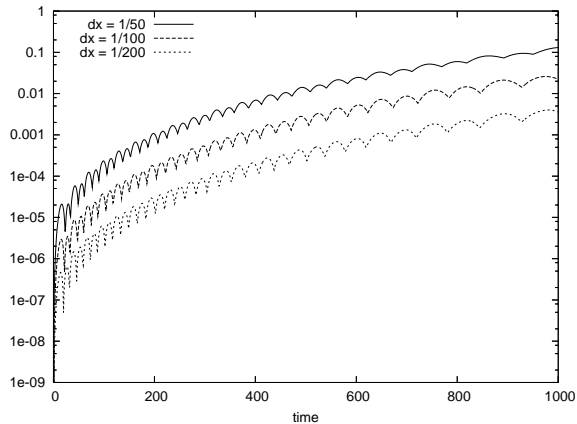}
\end{center}
\caption{The $\Theta$ quantity is plotted for three different resolutions
($dx=1/50, ~dx=1/100,~dx=1/200$) with the Q4 gauge and $\eta=1$, showing
a (second order) convergence to the exact zero solution.}
\label{gowdyconvergenceZo}
\end{figure}

The convergence of the solution for the ZQ4 system with $\eta=1$
is shown in Fig.~\ref{gowdyconvergenceZo}, where the $\Theta$
scalar is plotted for three different resolutions.

\section{Conclusions}

In this paper a large suite of hyperbolic gauge conditions has been studied with
some detail, pointing out their advantages and possible problems.

We have paid particular attention to conditions derived from
geometrical scalars devised in order to minimize spurious
coordinate effects. Our analysis reveal several important
consequences applicable to these conditions and also all related
ones (whose principal part coincide with those studied here, as
the gamma driver condition):

\begin{itemize}
  \item Minimization of these quantities leads to a characteristic structure that
  yields inflow modes near the black hole excision surfaces. This implies that some kind
  of boundary condition is needed. However, as mentioned this is a delicate issue as
  main and gauge variables are intertwined in the inflow modes.
  \item Related conditions, obtained by the addition of suitable advection terms to the equations, do
  resolve this issue but at a cost of bringing two more: the conditions do not necessarily minimize
  the sough-after scalars and there are surfaces where the system becomes weakly hyperbolic.
\end{itemize}
Notice that there is a way to avoid most these problems altogether
at the hyperbolic level by considering that a suitable first
integral of the conditions does exist (which could be ensured by
adding appropriate lower order terms to the equations)
\cite{LinSch03}. However, the resulting conditions need not
minimize the scalars and thus spurious coordinate effects might
very well remain.

As an alternative, a new coordinate condition, which has been
introduced very recently, is used as a gauge prescription for
Numerical Relativity applications. The main characteristic of this
gauge condition (Q4) is that tries to ``minimize" the deviation of
the time lines from the time-like (quasi) Killing vectors, if
there is one present on the space-time. The analogy with the 3D
minimal strain condition is pointed out, and the evolution
equations for the gauge quantities are written explicitly. The
full list of eigenvectors is given, showing how to pass from a
second order system (in space) to first order without changing the
structure of the eigenvectors. In order to enforce the desired
solution, some damping terms are included in the gauge equations,
which allow for two kind of interesting alternatives. The first
one corresponds to freezing all metric components and it can be
used when the space-time contains a (quasi) symmetry. The second
one does not attempt to minimized the rate change of the lapse,
but rather to drive it so that its rate of change is governed by
the trace of the distortion. This provides a less restrictive
alternative gauge condition for more general situations.

Finally, some numerical experiments have been performed in order
to check the properties of the Q4 gauge, the condition which
appears as the most promising one within generic hyperbolic
conditions derived in a geometrical way. First, with the robust
stability test, it has been shown that the evolution system leads
to solutions which are consistent with those of a well posed
problem even at the sonic points, where the system is weakly
hyperbolic just for some specific directions \footnote{Since weak
hyperbolicity only occurs for specific directions at these points,
while being strongly hyperbolic everywhere else, this issue is
often successfully dealt with some small amount of
dissipation~\cite{kreiss}}. The gauge waves test is also employed
to check the conditions, showing that the Q4 gauge (for the choice
$\eta=0$) indeed aligns the time lines with the time Killing
vector, thus leading to a stationary state. The Gowdy waves test
allows to further discriminate the effect of the damping
parameters, leading to either a singularity avoidant or to a lapse
freezing behavior (when the lapse is driven either by the
condition $Q \rightarrow tr Q$ or $Q \rightarrow 0$,
respectively).

\acknowledgements This research was supported in part by the NSF
under Grants No PHY0244335, PHY0244699, PHY0326311, PHY0302790 and
NASA-NAG-1430 to
Louisiana State University and the Horace Hearne Jr. Institute for
Theoretical Physics. The authors also acknowledge support from the
Spain's Ministry of Science and Education under project
FPA2004-03666. L.L. was partially supported by the
Alfred P. Sloan Foundation.

We thank C. Gundlach, H.O. Kreiss, J. Pullin, O. Sarbach and M. Tiglio, for
helpful discussions, suggestions and/or comments on the manuscript.

L.L. thanks the Isaac Newton Institute (Cambridge, UK) and the University
of Southampton (Southampton, UK) for hospitality during the last stages of this work.

\renewcommand{\theequation}{A.\arabic{equation}}
\setcounter{equation}{0}
\section*{Appendix: Sources of the Z4 evolution system}

\begin{eqnarray}\label{sources}
    S(K_{ij}) &=& - K_{ij}~{B_k}^k + K_{ik}~{B_j}^k +
    K_{jk}~{B_i}^k + \alpha\; \{\frac{1}{2}\; (1 + \xi)\; [-A_k\; {\Gamma^k}_{ij}
      + \frac{1}{2}(A_i\;D_j + A_j\;D_i)] \nonumber \\
      &+& \frac{1}{2}\; (1 - \xi)\; [A_k\; {D^k}_{ij}
    - \frac{1}{2} \{A_j\; (2\; E_i - D_i) + A_i\; (2\; E_j - D_j)  \}
    \nonumber \\
    &+& 2\; ({D_{ir}}^m ~ {D^{r}}_{mj}  + {D_{jr}}^m ~ {D^{r}}_{mi})
    - 2\; E_k\; ({D_{ij}}^k + {D_{ji}}^k)] \nonumber  \\
    &+& (D_k + A_k - 2\; Z_k) \; {\Gamma^k}_{ij}
    - {\Gamma^k}_{mj}\; {\Gamma^m}_{ki} - (A_i\; Z_j + A_j\; Z_i)
    - 2\; {K^k}_i\; K_{kj} \nonumber \\
    &+& (trK - 2\; \Theta)\; K_{ij}\}
    - 8\; \pi\; \alpha\; [S_{ij} - \frac{1}{2} \gamma_{ij} (- \tau +
    {S_k}^k)]
\\
   S(Z_i) &=& - Z_{i}~{B_k}^k + Z_{k}~{B_i}^k
   + \alpha\; [A_i\; (trK - 2\; \Theta) - A_k\; {K^k}_i
   - {K^k}_r\; {\Gamma^r}_{ki} + {K^k}_i\; (D_k - 2\; Z_k)]
   - 8\; \pi\; \alpha\; S_i
\\
   S(\Theta) &=& -\Theta~{B_k}^k
      + \frac{\alpha}{2}  [2\; A_k\; (D^k - E^k - 2\; Z^k)
      + {D_k}^{rs}\; {\Gamma^k}_{rs} \nonumber - D^k (D_k - 2\;
      Z_k) - {K^k}_r\; {K^r}_k \nonumber \\
      &+& trK\; (trK - 2\; \Theta)] - 8\; \pi\; \alpha\; \tau \nonumber \\
\end{eqnarray}

\end{document}